\documentclass[aps,prl,twocolumn,showpacs,amsmath,amssymb,superscriptaddress]{revtex4}
\usepackage{graphicx}

\begin{document}

\title{Fixation and consensus times on a network: a unified approach}

\author{G.\ J.\ Baxter}
\affiliation{School of Mathematics, Statistics and Computer Science, 
Victoria University of Wellington, PO Box 600, Wellington, New Zealand}

\author{R.\ A.\ Blythe} 
\affiliation{SUPA, School of Physics, University of Edinburgh, Mayfield Road, 
Edinburgh EH9 3JZ, UK}

\author{A.\ J.\ McKane}
\affiliation{Theory Group, School of Physics and Astronomy, University of 
Manchester, Manchester M13 9PL, UK}
 
\begin{abstract}
We investigate a set of stochastic models of biodiversity, population genetics,
language evolution and opinion dynamics on a network within a common framework.
Each node has a state, $0 < x_i < 1$, with interactions specified by strengths 
$m_{ij}$. For any set of $m_{ij}$ we derive an approximate expression for the 
mean time to reach fixation or consensus (all $x_i=0$ or $1$). Remarkably in a 
case relevant to language change this time is independent of the network 
structure. 
\end{abstract} 

%

\maketitle

Mathematical models predicting biological and social change are becoming 
increasingly commonplace, with the last few years having seen an explosion of 
activity among statistical physicists in cultural dynamics \cite{cas07}. One 
aspect of this work which is not widely appreciated is that several seemingly 
distinct phenomena can be described by very similar models: in some cases they 
can even be exactly mapped into each other \cite{bly07a}. Examples include 
biodiversity \cite{hub01}, population genetics \cite{cro70}, opinion dynamics 
\cite{cas07} and language change \cite{bax06}. The common thread is that 
objects that come in different variants are copied from one place (an 
``island'') to another according to some stochastic rule. If no new variants 
are created in the process (e.g., by mutation) and the number of objects does 
not grow without bound, one variant is eventually guaranteed to take over an 
entire population, or go to fixation in the genetics parlance. Changes in the 
network structure connecting different islands, and the stochastic rules used 
to choose the source and target islands, lead to a variety of scaling laws 
relating the number of islands and time to reach fixation (see, e.g., 
\cite{soo05,soo07,bly07}). In this Letter, we present a theoretical treatment 
of a very general stochastic copying model that includes many 
previously-studied cases and unifies the diverse fixation time results obtained
so far. We also discuss a mapping to a particle reaction system from which it
can be shown that our prediction for the fixation time, obtained by making 
various approximations, can in many cases be stated as a bound that simulations
show is often saturated. We also discuss a key application of our findings---to
a current theory of new-dialect formation \cite{tru04}.

To establish the basic features of the large class of models we consider, we 
describe a prominent special case, Hubbell's model of biodiversity and 
biogeography \cite{hub01}. Here there are only two islands: a metacommunity or 
mainland (island 1) and a local community (island 2). The objects are 
individuals which compete for a common resource (e.g.\ trees competing for 
space, sunlight and  nutrients \cite{con02}) and the variants are different 
species. At regular time intervals an individual in island 2 is picked at 
random to die and to be replaced by a copy of either (a) another individual 
picked at random from island 2, or (b) a \emph{migrant} from island 1. Process 
(b) is assumed to happen less frequently than (a), and as mentioned above, if 
(b) is absent then the final state of the system is one which contains 
individuals of only one species. The number of individuals on island 2 is a 
constant, $n$, and new species are created on island 1, but not island 2, by 
mutation-like events. This model is a \textit{neutral} theory; no one species 
is assumed to be ``fitter'' than another.  

\begin{figure}
\includegraphics[width=0.66\linewidth]{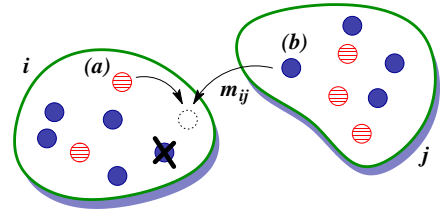}
\caption{\label{fig_dynamics} Stochastic copying dynamics between two islands 
$i$ and $j$ forming part of a larger network. After a death on $i$, an object 
is replaced either with a copy from (a) the same island $i$ or (b) a different 
island $j$ (rate $m_{ij}$).}
\end{figure}

The relationship between this model and simple neutral models of population 
genetics is well known \cite{hub01}: individuals are analogous to genes and 
species are types of genes (alleles). A more general model comprises a set of 
islands labeled $i=1,\ldots,N$ each of which contains $n$ genes (in reality, 
$n$ individuals each containing one copy of the gene of interest). For 
simplicity we assume that there is no mutation, so that no new alleles may be 
created. The alleles are labeled by $\alpha = 1,\ldots,M$. The dynamical 
processes, illustrated in Fig.~\ref{fig_dynamics} are as before: a death on 
island $i$ followed by (a) a birth on island $i$, or (b) a migrant offspring 
from island $j$ arriving on island $i$. There are various ways to parametrize 
these dynamics. As in \cite{bly07a}, we let the parent (copied object) be 
taken from island $j$ with probability $f_j$. In process (b) the probability 
the offspring (copy) lands on island $i$ is taken to be proportional to 
$m_{ij}$, which specifies a \emph{migration rate} within a standard 
continuous-time limit described in \cite{bly07a}, but whose details are 
unimportant here.  At any time $t$, the state of the system can be given in 
terms of the fraction $x_{i \alpha}(t)$ of genes on island $i$ that are allele 
$\alpha$.  Then, the islands can be represented as nodes on a network, which 
in the opinion dynamics and language change models are individuals $i$ with 
opinions or language variants $\alpha$ expressed with frequency $x_{i \alpha}$.
In these social contexts, fixation is sometimes called consensus.

Using the formalism first developed for these systems in population genetics 
\cite{cro70}, we describe the evolution in terms of a Fokker-Planck equation. 
Suppose, first of all, that there is only one island and no mutation. Then the
only dynamical process is \emph{random genetic drift} in which the frequencies 
$x_{\alpha} (t)$ diffuse on the interval $[0, 1]$ due to the random 
sampling in the death/birth process. For simplicity, we will assume that there
are only two alleles which have frequencies $x$ and $(1-x)$, so the state of 
the system is described by the single stochastic variable $x \in [0, 1]$. The 
probability that the system is in the state $x$ at time $t$, $P(x, t)$, 
satisfies the Fokker-Planck equation 
$\partial_{t} P(x, t) = \partial^{2}_{x} [ D(x) P(x, t)]$ where the diffusion 
constant is state dependent: $D(x) = x(1-x)/2$ \cite{cro70}. Moving to the 
case of $N$ islands, with migration rate $m_{ij}$ from $j$ to $i$, the 
Fokker-Planck equation for $P(x_{i}, t)$ now reads \cite{bly07a}
\begin{eqnarray}
\frac{\partial P}{\partial t} &=& \sum_{\langle i j \rangle} \left( m_{ij}
\frac{\partial }{\partial x_i} - m_{ji} \frac{\partial }{\partial x_j} \right)
\left[ \left( x_i - x_j \right) P \right] \nonumber \\
&+& \frac{1}{2} \sum^{N}_{i=1} f_{i} \frac{\partial^2 }{\partial x^{2}_{i}} 
\left[ x_{i} \left( 1 - x_{i} \right) P \right]\,,
\label{FP_eqn}    
\end{eqnarray}
where $\langle i j \rangle$ means sum over distinct pairs $i, j$. This may be 
generalized to $M > 2$ and to include mutation \cite{bly07a}, but 
Eq.~(\ref{FP_eqn}) will be sufficient for our purposes.

Analysis of Eq.~(\ref{FP_eqn}) is, on the face of it, a hopeless task since 
it has many degrees of freedom, $x_i$, interacting with arbitrary strengths 
$m_{ij}$.  However, much of the macroscopic dynamics is captured by the first 
and second moments of $x_{i}(t)$. The mean, 
$\alpha_{i} (t) = \langle x_{i} (t) \rangle$ can be found from 
Eq.~(\ref{FP_eqn}) to evolve according to
\begin{equation}
\frac{{\rm d}\alpha_{i}}{{\rm d}t} = \sum_{j \neq i} m_{ij} 
\left( \alpha_{j} - \alpha_{i} \right) \equiv \sum^{N}_{j=1} m_{ij} 
\alpha_{j}\,, 
\label{first_mom}
\end{equation}
where the equivalence holds if the diagonal elements $m_{ii}$ are defined to
be $-\sum_{j \neq i} m_{ij}$. This matrix has a zero eigenvalue, which we will 
assume is non-degenerate. The associated right eigenvector has all elements 
equal to one, and the left eigenvector we denote $Q_i$, so that
$\sum_{i} Q_{i} m_{ij} = 0$, and normalized such that 
$\sum_{i} Q_{i}=1$. Then we find from Eq. (\ref{first_mom}) that the ensemble 
(noise-history) average of the collective variable 
\begin{equation}
\xi (t) = \sum^{N}_{i=1} Q_{i} x_{i} (t)
\label{xi}
\end{equation}
is conserved by the dynamics. Decomposing $\alpha_{i} (t)$ in terms of its
right eigenvectors we see that it approaches a constant independent of $i$ as 
$t \to \infty$. Since all $x_{i} (t)$ tend to $0$ or $1$ as $t \to \infty$, 
this is the probability of the allele fixing. From (\ref{xi}) we see 
$\langle \xi (t) \rangle$ also approaches this value in this limit and since 
$\xi$ is conserved, the fixation probability is $\xi (0)$ \cite{suc05,soo05}. 

\begin{figure}
\includegraphics[width=\linewidth]{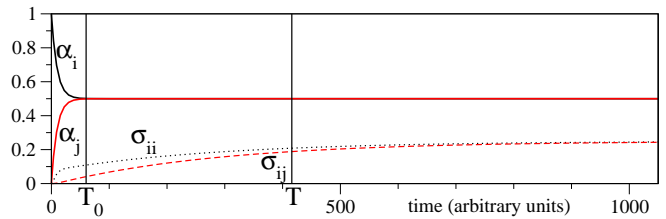}
\caption{\label{fig_approx} Numerical solution for means $\alpha_i$, 
$\alpha_j$ and (co)variances $\sigma_{ij}\equiv\beta_{ij}-\alpha_i \alpha_j$ 
on a fully-connected network of $N=20$ sites. Note the $\alpha$s converge at a 
time $T_0$ much less than the mean fixation time $T$, and at time 
$T_0$, $\sigma_{ij} \approx 0$ for $i\ne j$. These features become more 
pronounced as $N$ is increased.}
\end{figure}

Numerical studies suggest that convergence of the ensemble average 
$\alpha_{i} (t)$ to its asymptote $\xi(0)$ occurs on a much shorter timescale 
than the ultimate fixation of a variant, which in turn governs the rate of 
change of the second moments 
$\beta_{ij} (t) = \langle x_{i} (t) x_{j} (t) \rangle$, see 
Fig.~\ref{fig_approx}.  Eq.~(\ref{FP_eqn}) implies for the latter
\begin{equation}
\frac{{\rm d}\beta_{ij}}{{\rm d}t} = 
\sum_{k} m_{ik} \beta_{kj} + \sum_{\ell} m_{j \ell}
\beta_{i \ell} + \delta_{ij} f_{i} \left( \alpha_{i} - \beta_{ii} \right)
\label{second_mom}
\end{equation}
whilst the mean time to fixation, $T$, is given by the solution of a backward 
version of the Fokker-Planck equation (\ref{FP_eqn}) \cite{gar04}
\begin{eqnarray}
-1 &=& - \sum_{\langle i j \rangle} \left( x_i - x_j \right) \left( m_{ij}
\frac{\partial T}{\partial x_i} - m_{ji} 
\frac{\partial T}{\partial x_j} \right) \nonumber \\
&+& \frac{1}{2} \sum^{N}_{i=1} f_{i} 
\left[ x_{i} \left( 1 - x_{i} \right) \right] 
\frac{\partial^2 T}{\partial x^{2}_{i}}\,.
\label{BFP_eqn}    
\end{eqnarray}
The assumption that the time over which all the $\alpha_i$ converge, $T_0$, is 
much less than $T$ (see Fig.~\ref{fig_approx}) leads to the following 
approximate treatment of this equation.  We assume that $T$ depends \emph{only}
on the state of the system at time $T_0$, and principally through $\xi(0)$. 
Changing to the $\xi (0)$ variable using Eq. (\ref{xi}) we find that the first 
term on the right-hand side of Eq. (\ref{BFP_eqn}) vanishes \cite{soo05} giving
\begin{equation}
-2 = \sum^{N}_{i=1} f_{i} Q^{2}_{i} x_{i}(T_0) \left[ 1 - x_{i}(T_0) \right]
\frac{d^{2} T}{d\xi(0)^{2}}\,.
\label{T_eqn}
\end{equation}
This equation still depends on the variables $x_i$ at time $T_0$. We can 
estimate these by assuming that correlations between the nodes are absent, i.e.
$\beta_{ij} = \alpha_i \alpha_j = \xi (0)^2\ i \neq j$, and that the rate of 
change of the variance of $x_i$ is sufficiently slow that the time derivative 
in Eq. (\ref{second_mom}) when $i=j$ can be neglected (see again Fig.~2). Then 
$\beta_{ii}$ at time $T_0$ can be estimated from Eq. (\ref{second_mom}).

By replacing $x_i(1-x_i)$ in Eq. (\ref{T_eqn}) by 
$\alpha_i - \beta_{ii}$ at time $T_0$, we find 
$\left[ \xi(0)(1-\xi(0))\right] d^{2}T/d\xi(0)^{2} = - 2/r$ where 
\begin{equation}
r \approx \sum_{i} Q^{2}_{i} f_{i} \frac{2\sum_{j \neq i} m_{ij}}
{2\sum_{j \neq i} m_{ij} + f_{i}}\,.
\label{r}
\end{equation}
The mean fixation time is obtained by integrating the equation for 
$T(\xi(0))$, with the boundary conditions $T(0)=T(1)=0$ to give
\begin{equation}
T(\xi(0)) = - \frac{2}{r} \left[ \xi(0) \ln \xi(0) 
+ (1-\xi(0))\ln(1-\xi(0))\right]\,,
\label{fix_time}
\end{equation}
which in tandem with (\ref{r}) is our main result, derived for a large class 
of stochastic-copying processes on any network and arbitrary migration rates. 
It should be emphasized that the reduction from a system with $(M-1)$ degrees 
of freedom, to a system with only one degree of freedom is not imposed, but 
instead emerges from the dynamics.

For orientation and as a check of these results, we verify that specific 
choices of the parameters $f_i$ and $m_{ij}$ give expressions which have been 
previously obtained. As a first check, we examine the case of voter dynamics 
on heterogeneous graphs described in \cite{soo05,soo07}. The voter model proper
has each node chosen uniformly to be the recipient of a new opinion, hence all 
$f_i=1/N$. This opinion is selected randomly from the node's neighbors, giving 
$m_{ij} = A_{ij}/(2N k_i)$, where $A_{ij}=1$ if nodes $i,j$ are connected, $k_i$ 
is the degree of node $i$, $k_i = \sum_{j\ne i} A_{ij}$, and the factor of $2N$ is a consequence of the continuous-time limit described in \cite{bly07a}. To calculate $r$, one needs $Q_i$, the 
normalized zero left eigenvector of the matrix $m_{ij}$. One finds that this
eigenvector equals $Q_i = k_i / (N \bar{k})$, where $\bar{k}$ is the mean 
degree and is included to normalize the $Q_i$. Putting all this together we 
find from (\ref{r}) that $r =  \overline{k^2}/2(N\bar{k})^2$. This corresponds with the results of  \cite{soo05,soo07} after taking into account the different choice of time units made in those works. 

An invasion process, where the source node of an opinion is randomly selected in each timestep, was also considered in \cite{soo07}. This corresponds to the choice $m_{ij} = A_{ij}/(2N k_j)$, from which $Q_{i}$ may again be found. However, the resulting expression for $r$ via (\ref{r}) does not simplify unless one additionally assumes (as in \cite{soo05,soo07}) that node degrees are uncorrelated, $\overline{A_{ij}} = k_i k_j/(N \bar{k})$. Taking $f_i=1/N$ as before, one finds from (\ref{r}) the result of \cite{soo07} in the limit $N\to\infty$. Our treatment however also extends to networks with strong degree correlations, e.g., a star network that has a central node connected to all $N-1$ outer nodes, which in turn are connected only to the central node. For this network, (\ref{r}) predicts a fixation time proportional to $N^3$, confirmed by simulation data (not shown). By contrast, the approximation of \cite{soo07} predicts a fixation time that, in our time units, increases quadratically with $N$.

We also recover known results for link dynamics \cite{cas05,soo07}, where an 
opinion is copied in a randomly-chosen direction along a randomly-chosen link 
in each timestep. Here, we have $f_i=k_i/N$ and $m_{ij} = A_{ij} / (2N)$ which implies that $Q_i=1/N$. In this special case it has been noted that the mean time to fixation does not depend on the network structure \cite{cas05,soo07}. Furthermore, our general result (\ref{r}) includes the various scaling forms found in \cite{bly07}. 
 
The generality of our results can be tested more stringently by choosing $f_i$ and $m_{ij}$ from various random distributions in such a way that the numerators and denominators in (\ref{r}) are of similar magnitudes. Simulation results (not shown) on Erd\"os-R\'enyi random graphs of varying densities \cite{erd59} are consistent with Eqs.~(\ref{r}) and (\ref{fix_time}), except when $T$ and $T_0$ turn out to be of a similar order in $N$.

A concrete application of the model is to an evolutionary model of language 
change \cite{bax06}. In fact it was our experience in applying 
this model to the emergence of New Zealand English \cite{bax08} that alerted 
us to the possibility of a general analysis and to some surprising behavior 
for a subclass of models. As indicated earlier, in this particular application
the index $i$ labels speakers who converse and so affect each others grammar.
The probability that two speakers interact is given by a matrix $G_{ij}$; this
is a reflection of the topology of the network. Another factor $H_{ij}$ 
accounts for the weight that $i$ gives to the utterances of $j$. It is the 
product of the $G_{ij}$ and $H_{ij}$ that is equal to $m_{ij}$. While $G_{ij}$ 
is by definition symmetric, $H_{ij}$ need not be. In the application to the
formation of New Zealand English, Trudgill \cite{tru04} has postulated that the
large quantity of data that exists on the emergence on this English variant
\cite{gor04} may be explained by assuming that social factors which are modeled
by $H_{ij}$ are unimportant, and therefore this factor may be replaced by a 
constant. If this is so, the $m_{ij}$ is symmetric; if the model is not 
socially-neutral, then $m_{ij}$ will not in general be symmetric.

To test Trudgill's theory, we assume $m_{ij}$ is symmetric and using the 
results (\ref{r}) and (\ref{fix_time}) check if the mean time to fixation is
in accord with the data. A priori one would expect that the precise nature 
of the speaker network during the formation of New Zealand English would be 
required. However, remarkably, it turns out that in this case the time to 
fixation is completely independent of the network structure! This is because 
the symmetry of $m_{ij}$ implies that its left and right eigenvectors are 
proportional to one another, and so $Q_{i} = 1/N$. This together with 
$\sum_{j} m_{ij} \propto f_{i}$ \cite{bax06} implies a constant $r$ from Eq.
(\ref{r}), which is confirmed by simulation results shown in Fig.~3 (although 
see below for some caveats). To evaluate the plausibility of this 
socially-neutral model as a mechanism for new dialect formation requires in 
addition to this result detailed consideration of human memory lifetimes that 
are beyond the scope of this article and discussed elsewhere \cite{bax08}. 
Ultimately, we conclude that a purely neutral model is hard to reconcile with 
available empirical data.

\begin{figure}
\includegraphics[width=\linewidth]{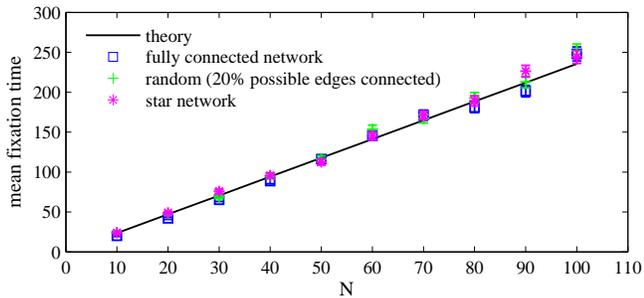}
\caption{\label{fig_lang} Fixation times \emph{per speaker} within the socially-neutral utterance 
selection model (defined in the text) on various networks with $N$ nodes (see 
legend).\vspace{-0.5cm}}
\end{figure}

We remark that our main result, Eq.~(\ref{r}), can also be obtained and 
interpreted within a \emph{backwards-time} formulation. Consider for example 
two objects of the same type; by asking which objects these were copied from 
at some earlier time, one can reconstruct their ancestral lineages which will 
hop between nodes as one goes back in time. Eventually, both objects will have 
been copied from the same `parent', causing merger of the lineages. This 
process is called the \textit{coalescent} in population genetics \cite{wak06}, 
but by viewing the lineages as particle trajectories, one can also recognize 
it as the $A+A\to A$ particle coalescence reaction of non-equilibrium 
statistical mechanics \cite{tau05} on a network.

In this picture, the quantity $r$ in Eq.~(\ref{r}) is the asymptotic rate at which the last two unreacted particles coalesce. The specific expression quoted above can be in fact obtained as a bound by means of a variational principle. Most usefully, this approach---whose details will be presented elsewhere \cite{new}---yields physical insight into when the approximations we have made to obtain (\ref{r}) are valid. The key assumption is that the last two unreacted particles must each typically have been able to explore a large number of the sites of the network before the final coalescence takes place. Then, we regard the particles as \emph{well-mixed}: the asymptotic probability of finding a particle on site $i$ relative to finding it on site $j$ is assumed to be independent of the location of the other particle \emph{unless} both particles are on the same site (since then they may react). In practice, we have found the presence of sufficient long-range connections in the network gives rise to a well-mixed two-particle state at late times in the particle-reaction process. On the other hand, a bottleneck that dramatically extends the time taken to reach one subset of sites from another would likely lead to this assumption of well-mixedness breaking down. Although (\ref{r}) would then cease to be valid, the variational approach of \cite{new} could, in principle, be refined to take such network structures into account, a task we leave for future work. Finally, we remark that the pivotal assumption $T_0 \ll T$ corresponds to taking the longest timescale in the backward-time dynamics to be that associated with the final coalescence reaction; that is, that the subsequent relaxation of the one particle state to its equilibrium occurs on a much shorter timescale. If the last two particles are well-mixed, we would expect this assumption to hold, since each particle would be close to the single-particle equilibrium state.

In summary, we have provided an approximate theory, validated numerically on a 
range of networks, for calculating the fixation time within a general model 
that has features shared by physical, biological and social systems. Since in 
reality specific applications will contain additional processes, it is of 
interest to extend the general approach we have described to cater for these. 
We have also briefly discussed how a particularly striking result---that 
fixation time is independent of network structure---is of direct relevance to 
current linguistic theory \cite{tru04,bax08} and hope that the trend whereby
mathematical results for formal models of social behavior are applied in 
empirical contexts will continue, and increase, in the future. 

RAB holds an RCUK Academic Fellowship. AJM wishes to thank the EPSRC (UK) for
financial support under grant GR/T11784.

\end{document}